\documentclass[]{elsart}
\usepackage[T1]{fontenc}
\usepackage[latin2]{inputenc}
\usepackage[polish,english]{babel}

\usepackage[centertags]{amsmath}
\usepackage{amsfonts}
\usepackage{newlfont}
\usepackage{epsfig}
\usepackage{amscd}

\newcommand{\RR}{{\mathbb R}}
\newcommand{\ZZ}{{\mathbb Z}}
\newcommand{\bN}{{\boldsymbol N}}
\newcommand{\bn}{{\boldsymbol n}}

\begin{document}

\begin{frontmatter}
\title{The integrable discretization of the Bianchi--Ernst system}
\author[N]{
M. Nieszporski
}
\author[D]{
A. Doliwa
}
\author[S]{
P.M. Santini
}
\address[N]{Instytut Fizyki Teoretycznej, Uniwersytet w Białymstoku,\\
ul. Lipowa 41, 15-424 Białystok, Poland\\
 e-mails: maciejun@fuw.edu.pl, maciejun@uwb.edu.pl
}
\address[D]{Instytut Fizyki Teoretycznej, Uniwersytet Warszawski,\\ 
ul. Ho{\.z}a 69, 00-681 Warszawa, Poland (corresponding author)\\
e-mail: doliwa@fuw.edu.pl}
\address[S]{Dipartimento di Fisica, Universit\`a di Roma ,,La Sapienza''\\ 
Istituto Nazionale di Fisica Nucleare, Sezione di Roma\\
P-le Aldo Moro 2, I--00185 Roma, Italy\\
e-mail: paolo.santini@roma1.infn.it}
\begin{abstract}
We present the constraint for the discrete Moutard equation which
gives the integrable discretization of the Bianchi--Ernst system.  
We also derive the discrete analogue of the Bianchi transformation between
solutions of such a
system (the Darboux-B\"{a}cklund transformation in soliton terminology).
We finally obtain  the superposition of discrete Bianchi transformations.
\end{abstract}
\end{frontmatter}

\section{Introduction}
\label{INTRO}
In his studies on isometric deformation of surfaces, Bianchi considered
\cite{Bianchi1} the Moutard equation
\cite{Moutard} 
\begin{equation}
\label{eq:Moutard}
\bN_{,uv}=f\bN,
\end{equation}
supplemented by the constraint
\begin{equation} \label{eq:NN-B}
\bN\cdot\bN =U(u)+V(v),
\end{equation}
where $\bN := (N_0,N_1,N_2)$,  "$\cdot$" denotes 
the scalar product (Bianchi considered the case $\epsilon=1$ only)
\begin{equation}
{\boldsymbol A}\cdot{\boldsymbol B}:=A_0B_0 + \epsilon (A_1B_1 + A_2
B_2), \qquad \epsilon = \pm 1
\end{equation} 
and $U(u)$, $V(v)$ are given funtions of single variables. 
In the considered reduction the function $f$ can be given in terms of $\bN$
and $U(u)$, $V(v)$ as follows
\begin{equation} \label{eq:BE-f}
f=-\frac{\bN_{,u}\cdot \bN_{,v}}{U(u)+V(v)}.
\end{equation}
Performing the following complex changes of the independent variables 
\begin{equation}
\label{eq:elipt}
u=\rho+i z, \qquad 
v=\rho-i z,
\end{equation}
and of the dependent ones
\begin{equation}
\label{eq:xi}
\xi=\frac{N_{1}+i N_{2}}{\sqrt{r}+N_{0}}, \qquad r=\bN\cdot\bN,
\end{equation}
equations \eqref{eq:Moutard}--\eqref{eq:NN-B} are transformed
into the following system
\begin{equation}
\label{eq:E}
\begin{array}{c}
(\xi \bar{\xi}+\epsilon) (\xi,_{\rho \rho}+\xi,_{zz}+
\frac{r,_{\rho}}{r} \xi,_{\rho}+
\frac{r,_z}{r} \xi,_{z})=2 \bar{\xi}  \left( (\xi,_{\rho})^2+ 
(\xi,_{z})^2 \right)\\
r,_{\rho \rho}+r,_{zz}=0, \qquad \epsilon = \pm 1.
\end{array}
\end{equation}
In the the case $\epsilon =-1$ the system \eqref{eq:E} was
considered by Ernst and describes
axisymmetric stationary vacuum Einstein fields \cite{Ernst}
as well as the interaction of gravitational waves \cite{A}.
Therefore we shall call the system  
\eqref{eq:Moutard}-\eqref{eq:NN-B} (or \eqref{eq:E})  the Bianchi--Ernst system.

The existence of  the
(Darboux--B\"{a}cklund) transformation between solutions of the 
Bianchi--Ernst system \eqref{eq:Moutard}-\eqref{eq:NN-B} 
(and therefore its integrability) was established in \cite{Bianchi1};
 for the interpretation of the Bianchi--Ernst system in the context of the 
modern soliton theory, see
\cite{Maison,Belinski,Neugebauer,Cenkl,Levi,Korotkin,Schief,Tafel}. 
In the simplest case $r=const$ the Bianchi--Ernst system reduces to the
$S^2$ chiral system equivalent to the sine-Gordon equation.
Therefore the Bianchi--Ernst system can  also be interpreted as a
non-isospectral integrable extension of a chiral model 
(a harmonic map or a nonlinear $\sigma$ model) 
see e.g. \cite{Mazur,Levi,Nieszporski} and references therein.

During the last few years the integrable discrete (difference) analogues of 
geometrically significant integrable differential equations have attracted
considerable attention \cite{B,C1,C2,D}. Indeed they represent the "building
blocks" of a new discipline, the integrable Discrete Geometry, and are also
potentially significant in a physical context.
 For example, the discrete analogue
of the sine-Gordon equation ($r=const$), known as the Hirota equation
\cite{Hirota}, not only describes the discrete analogue of pseudospherical
surfaces \cite{Sauer,BP1} and of Chebyshev nets
on  a sphere \cite{F}, but also turns out to be relevant in the analysis of some solvable
models of statistical mechanics and
quantum field theory (see also \cite{KBI,JM} for other examples).

In this Letter we introduce the integrable discrete analogue of 
the Bianchi--Ernst system \eqref{eq:Moutard}-\eqref{eq:NN-B}, namely the
system  
\begin{equation} \label{eq:BE-d}
\begin{array}{c}
\bN_{(12)} + \bN = F (\bN_{(1)} + \bN_{(2)}) ,\\
(\bN_{(12)} + \bN)\cdot(\bN_{(1)} + \bN_{(2)}) =U(m_1)+V(m_2),
\end{array}
\end{equation}
where subscripts in brackets denote shifts in the discrete variables $m_1$
and $m_2$ (for details see the next section). The above system
can be rewritten as the single nonlinear equation
\begin{equation} 
\bN_{(12)} + \bN = \frac{U(m_1)+V(m_2)}
{(\bN_{(1)} + \bN_{(2)})\cdot(\bN_{(1)} + \bN_{(2)})} 
(\bN_{(1)} + \bN_{(2)}) .
\end{equation}
We also establish, in the spirit
of Bianchi, the integrability of the system \eqref{eq:BE-d} obtaining:
i) the Darboux-type
transformation between solutions of the system and ii) the nonlinear 
superposition
principle for the solutions.
\begin{rem}
We consider in this paper the standard (in the continuous case) situation
$U(m_1)+V(m_2)>0$ only.
\end{rem}

The interesting geometric aspects  of the results of this Letter
are discussed in \cite{NieszporskiD,DSN},
 in the framework of
the theory of discrete asymptotic nets \cite{Doliwa,NieszporskiA}.
The relevance of the system \eqref{eq:BE-d} in the discrete theory of
gravity (see \cite{ReggeWilliams} and references therein) is an open and, in
our opinion, very important theoretical problem.

The Letter is organized as follows.
We first recall the necessary material concerning the discrete Moutard 
equation
and its Darboux-type (discrete analogue of the Moutard) transformation
\cite{Nimmo,Schief-DAS,Doliwa,NieszporskiA}. 
Then we introduce the proper analogue 
of the constraint \eqref{eq:NN-B}, which leads to the integrable reduction 
\eqref{eq:BE-d} of
the discrete Moutard equation. In the next step we study consequences of
such a constraint; in particular, we construct the corresponding reduction
of the discrete Moutard transformation. Finally we show the 
permutability of
the superposition of such transformations. 

{\em This work, fully inspired and motivated by the results presented in
\cite{Bianchi1}, is
dedicated to the memory of L. Bianchi, an outstanding precursor of the
modern theory of integrable systems.}

\section{The discrete Bianchi--Ernst system}
\label{sec:DBE}
Consider the mapping $\bN:\ZZ^2\to\RR^3$ satisfying the
discrete analogue of the Moutard equation \cite{Nimmo}
\begin{equation} \label{eq:Moutard-d}
\bN_{(12)} + \bN = F (\bN_{(1)} + \bN_{(2)}) ,
\end{equation}
where $F:\ZZ^2\to\RR$ is a given scalar function of the discrete variables
$(m_1,m_2)\in\ZZ^2$ and
$\bN_{(1)}(m_1,m_2) =\bN(m_1+1,m_2)$, 
$\bN_{(2)}(m_1,m_2) =\bN(m_1,m_2+1)$, 
$\bN_{(12)}(m_1,m_2) =\bN(m_1+1,m_2+1)$.
Given \cite{Schief-DAS}
a scalar solution $\Theta$ of equation \eqref{eq:Moutard-d}
\begin{equation} \label{eq:Moutard-d-Theta}
\Theta_{(12)} + \Theta = F (\Theta_{(1)} + \Theta_{(2)}) ,
\end{equation}
then the solution $\bN'$ of the system of equations
\begin{eqnarray}
\label{eq:dmt-1}
(\bN'_{(1)}\mp\bN )&=&\frac{\Theta}{\Theta_{(1)}}(\bN' \mp \bN_{(1)}) \\
\label{eq:dmt-2}
(\bN'_{(2)}\pm \bN )&=&\frac{\Theta}{\Theta_{(2)}}(\bN' \pm \bN_{(2)})
\end{eqnarray}
satisfies the Moutard equation \eqref{eq:Moutard-d} with transformed
potential
\begin{equation} \label{eq:dmt-F}
F'=\frac{\Theta_{(1)}\Theta_{(2)}}{\Theta\;\Theta_{(12)}}F.
\end{equation}
\begin{rem}
We consider \cite{Doliwa,NieszporskiA} two possibilities of signs in the Moutard
transformation in order: (i) to preserve the symmetry between the variables 
$m_1$ and $m_2$, and (ii) to reproduce the discrete Bianchi-Ernst equation 
in the superposition formula (see Theorem \ref{th:superposition} for
details).
\end{rem}
The following Lemma can be checked using equations 
\eqref{eq:Moutard-d}-\eqref{eq:dmt-2}.
\begin{lem} \label{lem:YY'ab}
Denote by
\begin{equation} \label{eq:Yab}
\begin{array}{ccl}
Y&:=&(\bN _{(12)} + \bN)\cdot (\bN _{(1)} + \bN _{(2)}),\\
Y'&:=&(\bN' _{(12)} + \bN')\cdot (\bN' _{(1)} + \bN' _{(2)}),\\
a&:=&(\bN' _{(1)} \mp \bN)\cdot (\bN' \mp \bN _{(1)}),\\
b&:=&(\bN' _{(2)} \pm \bN)\cdot (\bN' \pm \bN _{(2)}),
\end{array}
\end{equation}
where $\bN$ and $\bN'$ are connected by the discrete Moutard transformation
\eqref{eq:dmt-1}-\eqref{eq:dmt-2}, then the following identities holds
\begin{eqnarray}
Y-Y'&=&F\frac{\Theta_{(1)}+\Theta_{(2)}}{\Theta _{(12)}} (Y-a-b),\\
Y-a-b _{(1)}&=&F\frac{\Theta_{(1)}}{\Theta _{(12)}} (Y-a-b),\\
Y-a _{(2)}-b&=&F\frac{\Theta_{(2)}}{\Theta _{(12)}} (Y-a-b).
\end{eqnarray}
\end{lem}
The following important result is a straightforward consequence of
the above Lemma.
\begin{thm}
If $\bN$ and $\bN'$ are connected by the discrete Moutard transformation
\eqref{eq:dmt-1}-\eqref{eq:dmt-2}, then the condition
\begin{equation}
\label{eq:B-N-N'}
 (\bN_{(12)} + \bN )\cdot (\bN_{(1)} + \bN_{(2)}) =
 (\bN'_{(12)} + \bN')\cdot(\bN'_{(1)} + \bN'_{(2)})
\end{equation}
is equivalent to the following system of three equations
\begin{eqnarray}
\label{eq:B-N}
(\bN_{(12)} + \bN)\cdot(\bN_{(1)} + \bN_{(2)}) &=&U(m_1)+V(m_2),\\
\label{eq:B-N-N'1}
(\bN' _{(1)} \mp \bN)\cdot (\bN' \mp \bN _{(1)})&=&U(m_1)\mp k,\\
\label{eq:B-N-N'2}
(\bN' _{(2)} \pm \bN)\cdot (\bN' \pm \bN _{(2)})&=&V(m_2) \pm k,
\end{eqnarray}
where $U(m_1)$ and $V(m_2)$ are functions of single variables only 
and $k$ is a constant.
\end{thm}

\begin{rem}
Notice that the Moutard equation \eqref{eq:Moutard-d} subjected to 
the condition \eqref{eq:B-N} is exactly the discrete Bianchi--Ernst system
\eqref{eq:BE-d}, moreover the potential $F$ is of the form
\begin{equation} \label{eq:F-red}
F=\frac{U(m_1) + V(m_2)}{(\bN_{(1)} + \bN_{(2)})\cdot
(\bN_{(1)} + \bN_{(2)})} .
\end{equation}
\end{rem}

In this paper we are going to show the integrability of the reduction 
\eqref{eq:B-N} of the Moutard equation \eqref{eq:Moutard-d}. Let us first
derive some consequences of the condition \eqref{eq:B-N-N'}.

Denote by $\bn_0=\bN_{(12)} +\bN$ (we assume that
$\bn_0\cdot\bn_0>0$) and choose a pair of unit (of length $\epsilon$) 
vectors,
say  $\bn_1$ and $\bn_2$, such that $\{\bn_A\}_{A=0}^2$ form an 
orthogonal basis. Define functions $x^A$, $A=0,1,2$, as the coefficients
of the decomposition of the vector $\Theta (2\bN'\mp\bN _{(1)}\pm \bN _{(2)})$ in
that basis, i.e, 
\begin{equation}
\label{eq:xA}
\bN'= \frac{1}{2}\left(\pm \bN_{(1)} \mp \bN_{(2)} \right) +
\frac{x^A}{2\Theta}\bn_A,
\end{equation} 
where the summation convention holds. In virtue of the Moutard
transformation \eqref{eq:dmt-1}-\eqref{eq:dmt-2}, the above equation can 
be rewritten as
\begin{equation}
\label{eq:xA-2}
\Theta_{(1)} (\bN' _{(1)} \mp \bN)+ \Theta_{(2)}(\bN' _{(2)} \pm \bN)=
x^A \bn_A.
\end{equation}
The scalar multiplication of both sides of
equation \eqref{eq:xA-2} by $\bN _{(1)} +\bN _{(2)} $
gives, due to conditions \eqref{eq:B-N}-\eqref{eq:B-N-N'2},
\begin{equation} \label{eq:x0}
x^0 = \frac{\mp\Theta_{(1)} (U \mp k) \pm \Theta_{(2)}  (V \pm k)}{U+V}.
\end{equation}
Let us introduce the rotation coefficients $p_A^B$ and  $q_A^B$ by the unique
decompositions
\begin{equation} \label{eq:p-q}
\bn_A=p_A^B \bn_{B(1)}, \qquad \bn_A=q_A^B \bn_{B(2)}. 
\end{equation}
The compatibility condition for the above system reads
\begin{equation}
q_C^A p^B_{A(2)}  = p_C^A q^B_{A(1)} =:H_C^B.
\end{equation}
Inserting the expression \eqref{eq:xA} into the discrete Moutard 
transformation 
\eqref{eq:dmt-1}-\eqref{eq:dmt-2} and making use of the linear independence
of the vectors $\bn_A$, we obtain the following linear 
system of equations which the functions $x^A$ have to satisfy
\begin{equation} \label{eq:xA(i)}
\begin{array}{c}
x^A_{(1)} = p_B^A x^B \mp \frac{p_0^A}{F} \Theta
\pm (2 p_0^A-\frac{\delta_0^A}{F _{(1)} }) \Theta_{(1)}, \\
x^A _{(2)} =q_B^A x^B \pm \frac{q_0^A}{F} \Theta
\mp(2 q_0^A-\frac{\delta_0^A}{F _{(2)} }) \Theta_{(2)} ,
\end{array}
\end{equation}
where $\delta_B^A$ is the standard Kronecker symbol.

Equations \eqref{eq:xA(i)}, \eqref{eq:x0} and \eqref{eq:Moutard-d-Theta}
lead to the following linear system for the five unknowns 
$(\Theta,\Theta_{(1)},\Theta_{(2)},x^1,x^2)$

\begin{equation} \label{eq:lin-1}
\begin{pmatrix}
\Theta \\ \Theta_{(1)} \\ \Theta_{(2)} \\ x^1 \\ x^2
\end{pmatrix}_{(1)}=
\begin{pmatrix}
0&1&0&0&0 \\
\frac{Y_{(1)}\frac{p_0^0}{F}-b}{a_{(1)}}&
\frac{b F- Y_{(1)}(\frac{Y+b}{Y}p_0^0-\frac{1}{F _{(1)}})}{a_{(1)}}&
\frac{b}{a_{(1)}}(F-\frac{Y_{(1)}}{Y}p_0^0)&
\mp \frac{Y_{(1)}}{a_{(1)}}p_{1}^0&\mp\frac{Y_{(1)}}{a_{(1)}}p_{2}^0 \\
-1&F&F&0&0\\
\mp \frac{p_0^{1}}{F} & \pm\frac{Y+b}{Y}p_0^{1} &
\pm\frac{b}{Y}p_0^{1} & p_{1}^{1} & p_{2}^{1}\\
\mp\frac{p_0^{2}}{F} & \pm\frac{Y+b}{Y}p_0^{2} &
\pm\frac{b}{Y}p_0^{2} & p_{1}^{2} & p_{2}^{2}
\end{pmatrix}
\begin{pmatrix}
\Theta \\ \Theta_{(1)} \\ \Theta_{(2)} \\ x^1 \\ x^2
\end{pmatrix},
\end{equation}

\begin{equation} \label{eq:lin-2}
\begin{pmatrix}
\Theta \\ \Theta_{(1)} \\ \Theta_{(2)} \\ x^1 \\ x^2
\end{pmatrix}_{(2)}=
\begin{pmatrix}
0&0&1&0&0\\
-1&F&F&0&0\\
\frac{Y_{(2)}\frac{q_0^0}{F}-a}{b_{(2)}}&
\frac{a}{b_{(2)}}(F-\frac{Y_{(2)}}{Y}q_0^0)&
\frac{a F- Y_{(2)}(\frac{Y+a}{Y}q_0^0-\frac{1}{F _{(2)}})}{b_{(2)}}&
\pm\frac{Y_{(2)}}{b_{(2)}}q_{1}^0&\pm\frac{Y_{(2)}}{b_{(2)}}q_{2}^0
\\
\pm\frac{q_0^{1}}{F}&\mp\frac{a}{Y}q_0^{1}&\mp\frac{a+Y}{Y}q_0^{1}
&q_{1}^{1}&q_{2}^{1}\\
\pm\frac{q_0^{2}}{F}&\mp\frac{a}{Y}q_0^{2}&\mp\frac{a+Y}{Y}q_0^{2}
&q_{1}^{2}&q_{2}^{2}
\end{pmatrix}
\begin{pmatrix}
\Theta \\ \Theta_{(1)} \\ \Theta_{(2)} \\ x^1 \\ x^2
\end{pmatrix},
\end{equation}
where, according to the notation of Lemma \ref{lem:YY'ab} and the reduction 
under consideration, 
\begin{equation} \label{eq:Yab-red}
Y=U(m_1)+V(m_2), \qquad a=U(m_1) \mp k, \qquad b=V(m_2) \pm k.
\end{equation}

Finally, using formulas \eqref{eq:dmt-1}, \eqref{eq:B-N-N'1}, \eqref{eq:xA}
and the orthogonality of the basis $\{\bn_A\}_{A=0}^2$, one can derive the
algebraic constraint satisfied by the functions 
$\Theta$, $\Theta_{(1)}$, $\Theta_{(2)}$, $x^1$, $x^2$
\begin{equation} \label{eq:constraint}
\epsilon[(x^1)^2+(x^2)^2]+\frac{Y}{F}\Theta^2 +
FY\left( -\frac{a}{Y}\Theta_{(1)} +
\frac{b}{Y}\Theta_{(2)}\right)^2 -
2\Theta\left( a\Theta_{(1)}+b\Theta_{(2)} \right) =0;
\end{equation}
the same constraint one obtains using, instead of equations
\eqref{eq:dmt-1}, \eqref{eq:B-N-N'1}, equations \eqref{eq:dmt-2} and 
\eqref{eq:B-N-N'2}.

We are ready now to  present the main result of this paper. The following theorem gives
the discrete analogue of the Bianchi transformation
\cite{Bianchi1} between solutions of the discrete
Bianchi--Ernst system \eqref{eq:BE-d}. This transformation allows one
to find, using linear
steps only, new solutions of the discrete
Bianchi--Ernst system \eqref{eq:BE-d} from given ones.

\begin{thm} \label{th:dBtr}
Given a solution $\bN$ of the Bianchi--Ernst system \eqref{eq:BE-d} and 
the $\epsilon$-unit vectors $\bn_1$, $\bn_2$ (i.e.,
$\bn_1\cdot\bn_1=\bn_1\cdot\bn_1=\epsilon$) orthogonal to 
$\bN_{(12)}+\bN=:\bn_0$ and 
to each other, then
\begin{enumerate}
\item The linear system \eqref{eq:lin-1}-\eqref{eq:lin-2}, where $F$, $Y$, 
$a$,
$b$ are given by equations \eqref{eq:F-red}, \eqref{eq:Yab-red}, and
$p^A_B$, $q^A_B$ are given by \eqref{eq:p-q}, is compatible.\\
\item The solution $(\Theta,\Theta_{(1)},\Theta_{(2)},x^1,x^2)$ of the system 
\eqref{eq:lin-1}-\eqref{eq:lin-2} satisfies the constraint
\eqref{eq:constraint}, provided that such a constraint is satisfied at 
the initial point.\\  
\item Given the solution $(\Theta,\Theta_{(1)},\Theta_{(2)},x^1,x^2)$ of the 
system \eqref{eq:lin-1}-\eqref{eq:lin-2} satisfying the constraint
\eqref{eq:constraint}, then $\bN'$, constructed via equation \eqref{eq:xA}
with $x^0$ given by \eqref{eq:x0}, is a new solution of the discrete
Bianchi--Ernst system.
\end{enumerate}
\end{thm}
\begin{pf}
By direct, but tedious, verification. We list here only some useful
identities. To prove point 1. one makes use of the following identity
\begin{equation}\label{eq:n0form}
\frac{\bn_{0(12)}}{F_{(12)}} + \frac{\bn_{0}}{F} = \bn_{0(1)} + \bn_{0(2)},
\end{equation}
valid for any solution $\bN$ of the Moutard equation, from which one 
gets, via scalar multiplication by $\bn_{0(12)}$, $\bn_0$ and 
$\bn_{0(1)}-\bn_{0(2)}$,
\begin{equation}
\begin{array}{rcl}
\label{eq:n0form1}
\frac{F}{F_{(12)}}&=& F(p^0_{0 (2)}+q^0_{0 (1)})-H^0_0\\
H^0_A&=&\frac{-(U+V)\delta^0_A+p^0_A F_{(1)} (U_{(1)}+V)+
q^0_A F_{(2)} (U+V_{(2)})}{(U_{(1)}+V_{(2)})}\\
q^0_{0 (1)}&=&p^0_{0 (2)}+\frac{q^0_0 F_{(2)} (U+V_{(2)})-p^0_0 F_{(1)} 
(U_{(1)}+V)
+FF_{(1)} (U_{(1)}+V)-FF_{(2)} (U+V_{(2)})}{F(U_{(1)}+V_{(2)})}.
\end{array}
\end{equation}
To prove point 2. one makes use of a natural consequence of formulas 
\eqref{eq:p-q}
\begin{equation}
g_{AC}=p^B_A p^D_C g_{BD(1)}=q^B_A q^D_C g_{BD(2)}, \qquad
g_{AC}:=\text{diag}\left[(U+V)F,\epsilon,\epsilon \right].
\end{equation}
The proof of point 3. splits naturally into two parts. First, using point
1., we prove that such $\bN'$ is related to $\bN$ by the Moutard 
transformation \eqref{eq:dmt-1}-\eqref{eq:dmt-2} via the function
$\Theta$. Then, using the constraint \eqref{eq:constraint}, we immediately
check that $\bN'$ is subjected to 
\begin{equation}
(\bN'_{(12)} + \bN')\cdot(\bN'_{(1)} + \bN'_{(2)}) =U(m_1)+V(m_2). 
\qquad \qed
\end{equation}
\end{pf}
\begin{rem}
The parameter $k$, present in the linear system
\eqref{eq:lin-1}-\eqref{eq:lin-2}, is called the transformation parameter.
\end{rem}
\begin{rem}
The linear system \eqref{eq:lin-1}-\eqref{eq:lin-2} can also be interpreted
as a nonstandard Lax pair (zero curvature representation) of the discrete
Bianchi-Ernst system \eqref{eq:BE-d}, with spectral parameter $k$. A more
traditional Lax pair for system \eqref{eq:BE-d} will be presented in the
forthcoming paper \cite{DSN}.
\end{rem}
Finally, we present the superposition law associated with the
transformation described in Theorem \ref{th:dBtr}. 
The proof, omitted for the sake of  brevity (it will be published in 
\cite{DN}), consists in 
proving that the
discrete Bianchi constraint \eqref{eq:B-N} is  compatible 
with the superposition law of the discrete
Moutard  transformations \eqref{eq:dmt-1}-\eqref{eq:dmt-2}, given in 
\cite{NieszporskiA,Doliwa}.
\begin{thm} \label{th:superposition}
Given a solution $\bN$ of the Bianchi--Ernst system \eqref{eq:BE-d} and given
 two transforms of it (the upper index denotes the transformation!): 
the upper-sign transform $\bN^{(1)}$, with the transformation 
parameter $k^1$, and  the lower-sign transform
$\bN^{(2)}$, with the transformation parameter $k^2$. Then there exists the
unique solution $\bN^{(12)}$ of the Bianchi--Ernst system, given in algebraic
terms by
\begin{equation} \label{eq:BE-superp}
\bN^{(12)} = -\bN +
\frac{k^1+k^2}{(\bN^{(1)}+\bN^{(2)})\cdot(\bN^{(1)}+\bN^{(2)})}
(\bN^{(1)}+\bN^{(2)}),
\end{equation}
which is simultaneously the lower-sign transform of $\bN^{(1)}$, with 
the
transformation parameter $k^2$, and the upper-sign transform of 
$\bN^{(2)}$, with the transformation parameter $k^1$.
\end{thm}
\begin{rem}
Notice that the superposition formula \eqref{eq:BE-superp} for
the Bianchi--Ernst system \eqref{eq:BE-d} reproduces the 
Bianchi--Ernst system itself, after replacing the upper transformation
idices by the lower translation ones.
\end{rem}

\begin{rem}
Exactly like it was done in the continuous case \cite{Bianchi1,Nieszporski}
by taking $\bN \in \RR^n$ the above considerations can be generalized to 
the multicomponent discrete Bianchi--Ernst system.
\end{rem}

\end{document}